\documentclass[11pt]{article}

\usepackage[margin=1in]{geometry}

\usepackage{amsmath,amssymb,amsfonts,bm}
\usepackage[version=4]{mhchem}

\usepackage{graphicx}
\usepackage{float}
\usepackage{booktabs}
\usepackage{array}
\usepackage{authblk}
\usepackage{setspace}
\usepackage[dvipsnames]{xcolor}

\usepackage[numbers,sort&compress]{natbib}

\usepackage[T1]{fontenc}
\usepackage[utf8]{inputenc}
\usepackage{microtype}

\usepackage[colorlinks=true,
            citecolor=blue,
            linkcolor=blue,
            urlcolor=blue]{hyperref}

\title{Automated Many-Body Simulations of Strongly Correlated Systems Using a Correlation-Aware Agentic Framework
}

\author[1]{Tenghui Li}
\author[1]{Chong Sun\thanks{chongs0419@gmail.com}}

\affil[1]{Department of Chemistry and Chemical Biology, Rutgers University-New Brunswick,
Piscataway, NJ 08854, USA}

\date{}

\begin{document}

\maketitle

\begin{abstract}
We present CAFES, a correlation-aware agentic framework for electronic-structure simulations of strongly correlated systems. CAFES addresses two challenges: the fragmented software landscape for many-body calculations and the difficulty of selecting appropriate methods across diverse correlation regimes. It combines correlation diagnostics, adaptive method selection, and large language model (LLM) assistance for molecular, crystalline, and model-Hamiltonian systems. A study--task architecture separates study-level planning from task-level execution, while a curated scientific knowledge layer provides reusable guidance for method selection, workflow design, and result interpretation. We demonstrate CAFES through three research-level studies: calculating the low-lying electronic states of lutein using DMRG-CASSCF, probing phase competition in the extended honeycomb Hubbard model using DMET, and generating a quantum-chemical dataset with CCSD labels. These calculations demonstrate the potential of agentic workflows for strongly correlated electronic-structure problems, a regime that has received limited attention in existing agentic computational frameworks.

\end{abstract}

\section{Introduction}

Accurate simulation of strongly correlated systems remains a longstanding challenge in quantum chemistry~\cite{anisimov2010electronic, avella2012strongly}. Strong electron correlation can arise from distinct physical mechanisms and produce different numerical signatures. It may emerge when a chemical bond is stretched toward dissociation, when symmetry leads to degenerate or nearly degenerate transition-metal $3d$ orbitals, or when competing interactions frustrate the system.
These systems often require computationally demanding many-body approaches, including complete-active-space (CAS) methods~\cite{roos1980CASSCF}, the density matrix renormalization group (DMRG)~\cite{white1992dmrg,schollwock2005dmrg,chan2011density}, quantum Monte Carlo methods~\cite{hirsch1981efficient,becca2017quantum}, and quantum embedding methods~\cite{gutzwiller1965correlation,georges1996dmft,kotliar2006electronic,knizia2012dmet,wouters2016dmet,cui2020efficient,giordano2026ab}. Each method has its own systematic or statistical limitations, and no single method is uniformly reliable for all systems. Because experimental benchmarks are often limited for such systems, comparing results from multiple methods provides an important safeguard.
However, these methods typically require specialized expertise and method-specific software. Individual research groups also tend to focus on only one or two approaches, motivating large collaborative efforts such as the Simons Collaboration on the Many-Electron Problem~\cite{simonscollab2015hubbard, simonscollab2017hydrogen, zheng2017stripe}. Consequently, many-body calculations remain difficult to access for researchers without the necessary expertise.

Equally important is the ability to diagnose strong electron correlation. A computational study often involves multiple single-point calculations with varying geometries, interaction strengths, and electronic structures, so the resulting tasks can range from weakly to strongly correlated.
For instance, constructing a chemical reaction network requires sampling geometries across a potential energy surface (PES). At some geometries, near-degenerate orbitals or electronic configurations may make a multireference treatment necessary. Applying one level of theory uniformly across the entire study may therefore produce inaccurate results for strongly correlated geometries or incur unnecessary cost for weakly correlated ones. 
Systematic correlation diagnostics can address this problem by guiding the selection of an appropriate method for each calculation. In practice, however, identifying strong correlation often requires several diagnostic measures and their joint interpretation because strong correlation can arise from different electronic-structure features. Correlation diagnostics, such as coupled-cluster amplitude measures and natural-orbital occupation numbers, are themselves nontrivial to interpret and require expertise in computational chemistry~\cite{lee1989diagnostic, janssen1998diagostics, gordon1999natural, xu2024all}.

Recent developments in artificial intelligence (AI) agent frameworks provide a promising route toward automating many-body simulations. These frameworks combine large language models (LLMs) with tools, memory, and structured decision workflows to translate natural-language research questions into computational tasks and organize related calculations. Several platforms have been developed for general quantum-chemistry simulations~\cite{zou2025agente, perez2026agentequntur, ding2026automating, kieninger2026virtuallab, zhang2026qplanner, hu2025aitomia} and for computational workflows targeting specific systems~\cite{liu2026chemmate, liu2026masnet, lee2026simmof}. 
For strongly correlated simulations, the role of an agent extends beyond generating and executing individual tasks. An effective framework must connect method selection, diagnostic evidence, adaptive refinement, and comparison of related calculations, while preserving human oversight and computational traceability. Such a framework can make expert workflows easier to apply across systems and methods. Although recent work has begun to explore autonomous multireference calculations~\cite{lee2026can, wardzala2026multireference}, a unified workflow that combines these functions remains limited.

In this work, we introduce the Correlation-aware Agentic Framework for Electronic Structure (CAFES), built on the \textsc{PySCF} ecosystem and integrated with related packages. CAFES uses \textsc{PySCF} for general quantum chemistry simulations~\cite{sun2018pyscf, sun2020recent, sun2026python}, \textsc{Block2} for DMRG calculations~\cite{zhai2023block2}, \textsc{libDMET} for density matrix embedding theory (DMET) calculations~\cite{zheng2016hubbard2d, cui2020efficient, sun2020finite, knizia2012dmet, wouters2016dmet,  giordano2026ab}, and \textsc{fcDMFT} for dynamical mean-field theory (DMFT) calculations~\cite{zhu2019coupled, zhu2020efficient, zhu2021ab, georges1996dmft, kotliar2006electronic}.
To address the methodological diversity and repeated computational decisions involved in studies of strongly correlated systems, CAFES provides a unified framework spanning molecules, crystalline systems, and model Hamiltonians. Within this framework, related calculations are organized into a study, capturing relationships among individual tasks and enabling coordinated execution. As the study progresses, correlation diagnostics probe the character of electron correlation and identify potential limitations of the current  treatment, while a curated knowledge layer supports their interpretation and guides method refinement and chemically informed active-space selection. Throughout this design, CAFES separates LLM-assisted planning and interpretation from numerical execution. 

CAFES provides access to its computational capabilities through a web interface, command-line tools, Python application programming interfaces (APIs), and a Model Context Protocol (MCP) interface for integration with external agents. Within a common workflow, it supports model Hamiltonian construction, diagnostic-guided method refinement, and chemically informed active-space selection and review. CAFES carries these scientific choices through to numerical execution and analysis, allowing results to be assessed across a study and individual calculations to be revisited when further refinement is needed.

We illustrate these capabilities through three research-level applications: DMRG-CASSCF calculations on the lutein molecule with active-space selection, DMET  studies of the honeycomb Hubbard model quantum phases, and molecular dataset generation with CCSD labels. Across these applications, CAFES coordinates successive calculations while preserving the links between methodological choices, diagnostic evidence, and numerical results. By reducing routine coordination and retaining the basis for computational decisions, CAFES can make advanced many-body methods easier to use and support reproducible research, with scientific judgment remaining under researcher oversight. 

The remainder of this article is organized as follows. Section~\ref{sec:framework} describes the design and architecture of CAFES. Section~\ref{sec:features} presents its main computational features, and Section~\ref{sec:results} reports three computational studies performed with the framework. Finally, Section~\ref{sec:conclusion} discusses the scientific role, numerical reliability, and generality of agentic workflows and summarizes the main conclusions.

\begin{figure*}[t]
    \centering
    \includegraphics[width=0.9\linewidth]{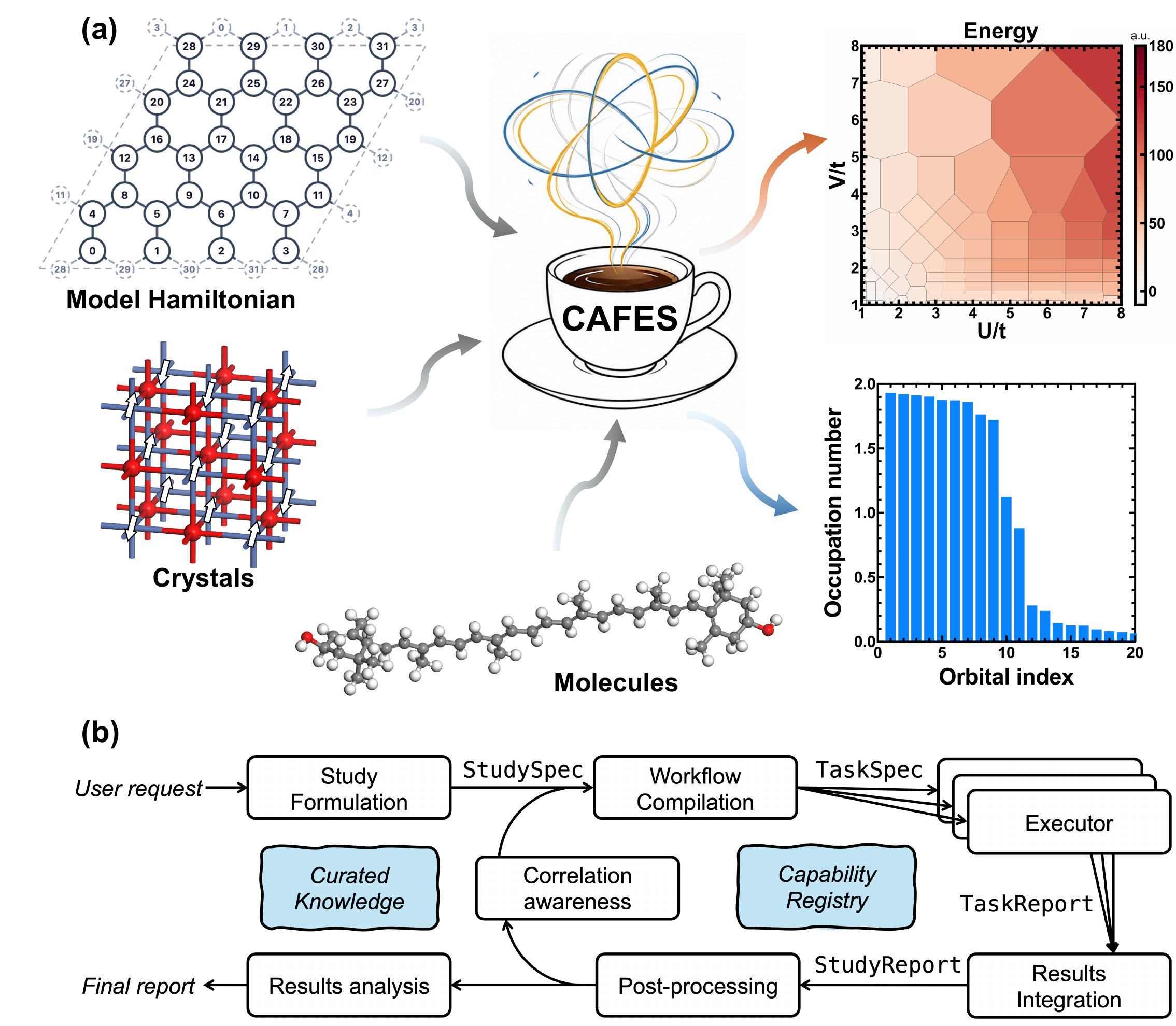}
    \caption{(a) Overview of the CAFES framework. (b) The CAFES workflow with a study-task organization. 
    }
    \label{fig:overview}
\end{figure*}

\begin{figure*}[t]
    \centering
    \includegraphics[width=0.9\linewidth]{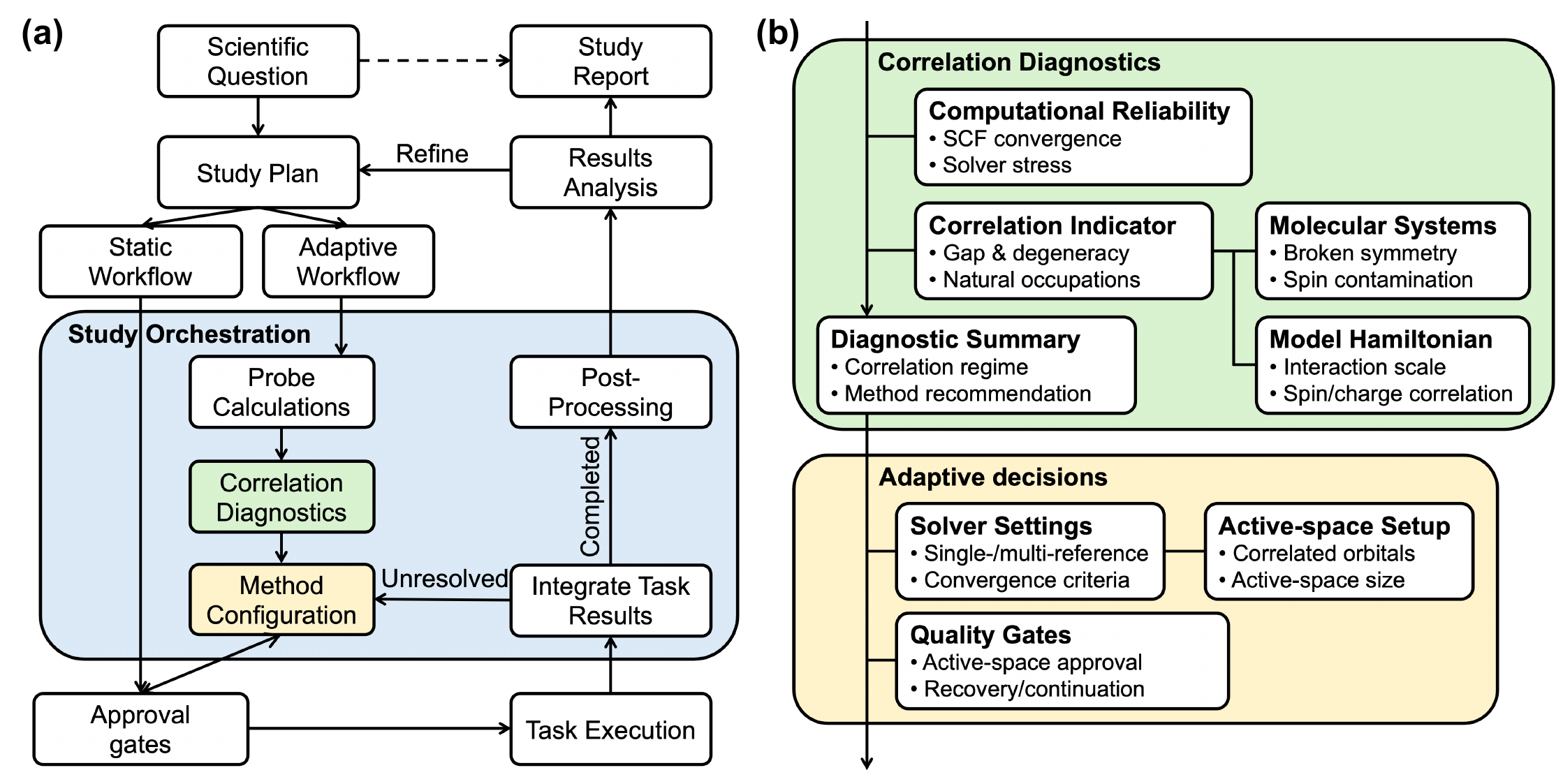}
    \caption{Overview of study orchestration and correlation-aware decision making in CAFES. (a) Organization of static and adaptive studies. (b) Electronic-structure indicators and computational-reliability checks, and their role in guiding the choice of numerical methods, solver settings, and active spaces.}
    \label{fig:workflow}
\end{figure*}

\section{Framework and design\label{sec:framework}}
We adopt a \textit{study--task} framework to organize computational workflows, which separates the scientific reasoning of a study from the execution of individual tasks. This separation is particularly important for strongly correlated systems, where related calculations may require different physical models, levels of theory, and numerical methods, in contrast to conventional workflows that often rely on a fixed computational protocol. CAFES is therefore designed to accommodate such methodological heterogeneity within a common structure, allowing individual task implementations to remain system and method specific.



\subsection{Design principles}

\textbf{Study--Task organization.}
A computational investigation is organized into \textit{studies} and \textit{tasks} with distinct responsibilities, as illustrated in Fig.~\ref{fig:overview} (b). The study layer coordinates a set of related tasks within a broader scientific investigation. It relates individual tasks to the scientific objective, collects and analyzes task reports, and uses the accumulated results to determine subsequent calculations. A task represents an individual calculation with a specified system setup, computational objective, and numerical method. This separation allows computational details to vary across tasks while maintaining a common scientific context. It also allows the same calculation procedure to be executed either independently or as part of a larger study. The study- and task-level descriptions are recorded in structured specifications, i.e., \texttt{StudySpec} and \texttt{TaskSpec}, which make the study objectives and individual calculation settings explicit.

\noindent\textbf{Capabilities and knowledge.} 
The study--task organization defines how calculations are coordinated, but it does not determine which calculations are scientifically appropriate for a given system or objective. Choosing an appropriate method therefore requires knowledge of both the capabilities of CAFES and the scientific conditions under which each method is applicable. CAFES addresses this through a unified capability registry and a curated knowledge layer, shown in Fig.~\ref{fig:overview} (b). The registry specifies the supported calculations and their conditions of use, while the wiki provides scientific context for these methods, including their assumptions, limitations, and possible combinations. Together, these components provide LLMs with the scientific context needed for planning while keeping executable capabilities and computational constraints explicitly defined.


\noindent\textbf{Separation of reasoning and execution.} 
Planning a study requires interpreting the scientific question and mapping it to appropriate capabilities. CAFES uses LLMs to assist this process by selecting and combining available capabilities into a computational workflow. These choices must nevertheless be validated before numerical execution, since a plausible plan may omit required inputs or specify incompatible settings. CAFES therefore validates inputs, method compatibility, and required approvals before execution. These checks, together with the subsequent numerical execution, follow predefined rules without LLM involvement. This design allows LLMs to guide scientific planning while keeping the conditions and procedures governing execution explicit and deterministic.

Since numerical execution is governed by the registry and deterministic qualification rather than by the language model, CAFES is model-agnostic: any LLM accessible through a compatible interface can serve as the reasoning component without changes to the workflow or backend. 

\subsection{Structured representation}

CAFES makes many-body simulations accessible by allowing users to express scientific goals in natural language. Numerical execution, however, requires explicit and validated inputs, whereas natural-language requests may omit essential settings such as charge, spin, or boundary conditions. CAFES bridges this gap by converting user requests into structured specifications and recording calculation outcomes as structured reports. These representations are used throughout the study--task workflow, connecting natural-language input to computational backends and providing a consistent basis for validation and result analysis. A common format, extensible with system-specific fields, supports molecular, crystalline, and model-Hamiltonian calculations.

CAFES defines four data structures, \texttt{StudySpec}, \texttt{TaskSpec}, \texttt{TaskReport}, and \texttt{StudyReport}, as shown in Fig.~\ref{fig:overview}(b). A \texttt{StudySpec} describes a multi-task scientific objective, including the base calculation, task-generation procedure, shared settings, and requested analysis. CAFES expands the \texttt{StudySpec} into explicit tasks, each represented by a \texttt{TaskSpec} containing the normalized system description and numerical settings required for validation and workflow compilation.

Execution produces a \texttt{TaskReport} containing the method configuration, numerical results, diagnostics, and validation information. After the tasks are executed, their reports are collected into a \texttt{StudyReport}, which records adaptive decisions and post-processing outputs and links each result to its parent study. Large output files, checkpoints, and numerical arrays are stored externally and referenced from the reports rather than copied into the conversational context. This preserves access to the underlying evidence while allowing LLMs to work with concise summaries during analysis.

\subsection{Registry and task workflow}

\textbf{Capability registry.}
CAFES maintains a unified capability registry for workflow planning, validation, and execution. Each supported operation is registered as a module with a defined role, interface, typed parameters and observables, compatibility conditions, quality gates, and runtime binding. Table S1 summarizes the capability registry, and representative entries are provided in the Supplementary Material. The same registry is used by the user interfaces, study planner, and numerical backend, avoiding separate and potentially inconsistent capability lists.
The registry also describes dependencies between operations and the conditions under which they can be composed. An operation may require outputs from an earlier stage, a particular system type, or explicit user approval. For example, a CAS-based calculation can invoke a target solver only after an active space has been prepared and approved. These constraints allow CAFES to determine whether supported operations can be combined before execution.

\noindent\textbf{Workflow compilation.} CAFES compiles each qualified task into a workflow of registered modules. The compiler maps the abstract task specification to concrete execution steps, using a common sequence of input preparation, validation, execution, result collection, and report generation. It adds system- and method-specific modules as needed, including model construction, orbital localization, active-space preparation, and embedding setup.

The task workflow is implemented as a LangGraph state graph~\cite{langgraph2024github}. The graph defines permitted transitions between stages, while the compiler selects modules, resolves dependencies, and determines their execution order. This process also identifies auxiliary stages required by the selected treatment. For example, orbital localization may prepare inputs for correlated or embedding calculations, while active-space preparation produces an \texttt{ActiveSpaceAudit} record for review before a CAS solver is invoked. The resulting workflow therefore has a common structure while remaining specific to the task.

\noindent\textbf{Qualification.} Qualification checks whether a study or task is complete and internally consistent with the capability registry. Compilation is request-driven: it starts from the scientific specification and selects the modules needed for execution. Qualification is requirement-driven: it checks whether the selected modules have their required inputs, satisfy compatibility conditions, and pass applicable gates. Both processes use explicit registry constraints and program logic rather than LLM judgment.

Study-level qualification checks whether the generated tasks satisfy the original \texttt{StudySpec}, including the system type, task-generation procedure, method settings, and requested comparisons. Task-level qualification checks each \texttt{TaskSpec}, including its method, inputs, and compatibility with registered module requirements. Missing information or inconsistencies produce structured feedback for clarification or revision. This prevents CAFES from silently replacing the requested calculation with a different one. Both levels are necessary because a valid study specification does not guarantee that every generated task is executable; changes in geometry, electron count, or solver settings can create task-specific incompatibilities.

\subsection{Study orchestration}

The study layer builds on task-level mechanisms to coordinate related calculations within a single scientific investigation. It maintains task-generation rules, completed reports, refinements, and review decisions in one study context, allowing results from different stages to be analyzed together.

\noindent\textbf{Task generation.} CAFES supports static and adaptive study modes. Static studies are appropriate when the computational method is known in advance, such as when applying the same method across a series of geometries, model-Hamiltonian parameters, or related systems. Adaptive studies are intended for cases in which the appropriate treatment depends on results obtained during the study. For example, in potential energy surface (PES) exploration, the importance of strong electron correlation may vary across geometries. CAFES can therefore begin with probe calculations, use their diagnostic information to assess the selected geometries, and then choose suitable methods or numerical settings for subsequent tasks.

\noindent\textbf{Adaptive refinement.} Adaptive refinement uses evidence from completed tasks to determine whether additional calculations are needed and which tasks should be reconsidered. This evidence may come from probe calculations, execution outcomes, or post-processing analysis. Probe calculations may indicate strong electron correlation, execution outcomes may identify failed or unconverged calculations, and post-processing may reveal results that are inconsistent with related tasks. These observations provide the basis for selecting tasks for refinement.

For each selected task, CAFES proposes a revised task specification. Depending on the available evidence, the refinement may adjust numerical settings, select a different method, or request an active-space review. LLMs may assist in interpreting the evidence, but every proposed change is recorded explicitly in a revised \texttt{TaskSpec} that passes the same qualification checks as the initial tasks.

\noindent\textbf{Human-in-the-loop.} Some study decisions may require substantial computational resources or may materially affect the scientific interpretation. Such decisions are therefore not made solely by the automated workflow; they require approval at the applicable gates. Depending on the workflow, these gates may apply to steps such as geometry generation, active-space selection, and other computationally expensive or scientifically consequential choices. CAFES records each approval together with the corresponding specification, allowing subsequent analysis to refer to an explicit review record rather than to the conversation history.

\noindent\textbf{Execution and recovery.} After qualification and compilation, CAFES generates and archives the numerical input files and scripts, preserving the procedure used for each calculation. A study may involve many computationally demanding calculations. CAFES supports local execution and cluster scheduling, including remote Slurm submission. CAFES monitors job status and numerical progress, with execution displayed in the web interface. 

A completed scheduler job does not guarantee numerical success. CAFES therefore examines output files, convergence behavior, and diagnostic information before assigning the task outcome. If a calculation fails or does not converge, recovery strategies may adjust numerical settings within predefined limits. If recovery is unavailable or unsuccessful, CAFES records the failure rather than concealing it by switching to a different scientific treatment.

When further calculations are needed, CAFES presents proposed refinements or recalculations, together with their supporting evidence, for user approval. Once approved, only the affected tasks are executed, while results from unaffected tasks are retained. Each calculation attempt receives a separate run identifier, and its inputs, outputs, and diagnostic records are preserved without overwriting those from earlier attempts. Results from subsequent calculations are incorporated into the same \texttt{StudyReport}, providing evidence for further refinement or final analysis.

\subsection{Curated knowledge layer}

In addition to structured specifications and registry-defined workflows, CAFES maintains an LLM-accessible scientific wiki to support reasoning throughout a study. This is important for strongly correlated systems, where method selection, interpretation of diagnostics, and result analysis often require expert knowledge that cannot be fully represented by fixed rules or permitted-parameter lists. The wiki contains method descriptions, workflow guidance, diagnostic explanations, known failure patterns, and implementation limits.
Relevant entries can be retrieved during request formulation, workflow planning, and result interpretation. They guide the selection and combination of capabilities, explain reported quantities, and clarify their scientific meaning and limitations. For example, in planning the continuation of the lutein study, the relevant entries explain that reusing a converged matrix product states (MPS) requires retaining the orbital basis and ordering in which it was obtained. Accordingly, the optimized orbitals were transferred together with the MPS between successive calculations.

CAFES and users can record observations from calculations, failure analyses, and researcher feedback, together with their context. LLMs can organize these records into proposed wiki additions, which are reviewed before inclusion. For instance, failures in the honeycomb DMET scan were recorded according to the stage at which they occurred: the mean-field calculation, bath construction, impurity solution, or the update of the correlation potential. These records were organized into a proposed wiki entry that relates each failure stage to recovery options, such as switching to the alternative broken-symmetry initial density, increasing the maximum number of iterations, or continuation from a converged neighboring parameter point. The wiki can therefore evolve as experience accumulates, making knowledge from earlier studies available for subsequent planning.

\section{Program Features\label{sec:features}}

\subsection{User interfaces}

CAFES integrates study preparation, execution monitoring, and result analysis through three interfaces: a web interface for interactive study preparation and result inspection; command-line tools and Python APIs for scripted calculations; and an MCP interface for access by external agents. All three use the common study--task representations described above, supporting consistent calculation settings and reproducible, traceable workflows.

\noindent\textbf{Web interface.} The web interface comprises the Study Planner, Calculation Assistant, and Model Hamiltonian Builder (Figs. S1 and S2). The Study Planner displays the tasks generated for a study, including their inputs, methods, execution status, and results, allowing users to review and follow the study as it progresses.
The Calculation Assistant converts natural-language requests into executable tasks for molecular, crystalline, and model-Hamiltonian systems. It provides validation feedback on missing inputs and incompatible settings and presents relevant evidence for human review. Users can inspect and revise the proposed workflow and numerical parameters before execution.
The Model Hamiltonian Builder provides graphical tools for constructing Hubbard, Hubbard--Holstein, and one-dimensional Su--Schrieffer--Heeger (SSH) models on various lattice geometries, including square, honeycomb, kagome, and square-octagon lattices~\cite{hubbard1963model, holstein1959studies, su1979ssh}. Users can select a lattice template and edit site positions, bond connections, hopping amplitudes, interaction strengths, and electron filling. The resulting model can be sent to the Calculation Assistant for an individual calculation or to the Study Planner for a related set of calculations.

\noindent\textbf{Command-line and Python interfaces.} CAFES provides command-line tools and Python APIs for repeated calculations and integration with existing computational workflows. The command-line tools accept saved task or study specifications, while the Python APIs allow specifications to be constructed and processed within custom code. Both interfaces use a common specification format across backends and provide pre-execution validation, provenance tracking, and restartable study-level orchestration without LLM involvement. For example, the honeycomb scan in Section 4.2 is defined by a single \texttt{StudySpec} rather than separate scripts for setup, submission, and result collection.

\noindent\textbf{MCP interface.} CAFES exposes its calculation and study services through the Model Context Protocol (MCP), allowing external AI assistants and agents to use these services in their own workflows. An external agent can consult the capability registry and curated knowledge, prepare a request, and submit it to CAFES for validation and execution. Requests, execution information, and results are exchanged through the shared structured representations. Validated execution and recorded results provide reproducibility and traceability, while grounding agent interpretations in calculation outputs helps limit unsupported numerical claims. Reusing established workflows and compact representations can also reduce the amount of code and raw output that an agent must generate or process.

\subsection{Electronic structure simulations}
CAFES supports electronic structure simulations of molecules, crystals, and model Hamiltonians using methods from the \textsc{PySCF} ecosystem, as described below. These methods are registered in the capability registry, allowing them to be selected and combined according to the physical system and the desired treatment of electron correlation.
 
\noindent\textbf{Molecular systems.} Electron correlation in molecular systems can change with geometry, charge, and spin state. For example, in chemical reactions, bond stretching can increase the importance of configurations that are negligible near equilibrium, making a single-reference method suitable at equilibrium inadequate. CAFES addresses this by generating tasks along a specified reaction coordinate while keeping other settings fixed. Initial calculations can use a cheap method across the scan, with correlation diagnostics identifying geometries that require further investigation. 

CAFES supports single-reference methods, including M{\o}ller--Plesset perturbation theory and coupled cluster methods \cite{moller1934note, purvis1982ccsd, bartlett2007coupled}, as well as multireference methods such as complete active space self-consistent field (CASSCF)~\cite{roos1980CASSCF} and $n$-electron valence state perturbation theory (NEVPT)~\cite{angeli2001introduction, angeli2001NEVPT2}. These methods allow studies to compare the effects of different correlation treatments on calculated dissociation curves. For active spaces too large for conventional configuration-interaction solvers, the \textsc{Block2} interface provides DMRG solvers~\cite{zhai2023block2} for supported active space calculations.

\noindent\textbf{Crystalline systems.} Crystals are modeled as periodic systems composed of repeating units, so their electronic structure is often described in reciprocal ($k$) space. CAFES integrates \textsc{PySCF}'s periodic-system capabilities, including $k$-point sampling, basis-set and pseudopotential selection, and density fitting. Calculations begin with a periodic Hartree--Fock (HF) or DFT treatment, which provides mean-field properties such as the band structure.

This mean-field reference can then be used for correlated calculations with DMET or DMFT, supported through the \textsc{libDMET} and \textsc{fcDMFT} packages, respectively~\cite{zheng2016hubbard2d,cui2020efficient,sun2020finite,zhu2019coupled,zhu2020efficient,zhu2021ab}. DMET is typically less computationally demanding and is suitable for static properties, including local magnetism, spin and charge correlations, and double occupancy. It also supports systematic extrapolation toward the thermodynamic limit. DMFT is used for dynamical quantities such as Green's and spectral functions.

A unified workflow allows the mean-field reference and orbital-localization procedures to be shared between DMET and DMFT, reducing redundant calculations and simplifying maintenance. The web interface further allows users to configure settings such as the fragmentation, impurity solver, and bath construction, reducing the need to specify these details manually in input scripts.

\noindent\textbf{Model Hamiltonians.} Model Hamiltonians provide simplified descriptions of the relevant low-energy degrees of freedom in realistic systems. They can be derived through downfolding or constructed phenomenologically. By reducing the number of degrees of freedom, they make otherwise prohibitive calculations tractable while retaining the relevant low-energy physics. They also enable systematic studies of the effects of interactions, filling, and connectivity. This flexibility supports the exploration of phases across related systems and informs the design of functional materials. Therefore, model-Hamiltonian studies are an important area of computational research in their own right.

CAFES supports model-Hamiltonian calculations using methods registered for the selected model and system representation. In particular, DMRG, DMFT, and DMET are commonly used for strongly correlated model systems. The Model Hamiltonian Builder, introduced earlier, provides an intuitive graphical web interface for constructing models, visualizing their connectivity, and adjusting their parameters. It allows users to construct model Hamiltonians without manually writing Hamiltonian terms in package-specific formats.

\subsection{Correlation diagnostics}

``Strong correlation'' is an umbrella term for systems with different physical origins and computational manifestations. In quantum chemistry, it often refers to static correlation arising from near-degenerate electronic configurations that are poorly described by a single Hartree--Fock reference. In condensed matter systems, strong correlation may arise from electron localization, strong on-site interactions, long-range fluctuations, or geometrical frustration. Identifying the nature of the correlation is important for selecting appropriate computational methods and interpreting their results. CAFES therefore uses on-the-fly diagnostics to identify limitations of the current treatment and guide method selection and refinement.

\noindent\textbf{Correlation diagnostics.}
CAFES groups correlation diagnostics into two categories: electronic-structure indicators and calculation-reliability checks. Electronic-structure indicators include small orbital gaps, near-degeneracies, and fractional natural-orbital occupations. Because no single indicator is sufficient, CAFES supplements these general measures with system- and method-specific diagnostics. For molecular systems, these include spin contamination and symmetry breaking. Mean-field and post-Hartree--Fock probe calculations provide additional information through orbital occupations and wavefunction amplitudes~\cite{lee1989diagnostic,janssen1998diagostics}.

Reliability checks assess whether the underlying calculations are sufficiently converged and stable for these indicators to be interpreted. CAFES therefore evaluates convergence and SCF-stability results alongside the electronic-structure diagnostics. It scores the two categories separately and combines them using predefined weights to obtain an overall diagnostic score (Eq. S1 and Table S2). These weights could in future be learned and updated as computational evidence and scientific experience accumulate, allowing CAFES to identify which indicators are most informative for a given system or method and to apply that knowledge in later studies.

\noindent\textbf{Method advisor.} The method advisor uses these diagnostic assessments to recommend subsequent calculations according to rules documented in the scientific wiki. For molecular systems, the default policy assigns single-reference coupled-cluster treatments when correlation indicators are weak and CAS-based treatments when they indicate substantial multireference character. Intermediate cases retain the single-reference treatment but are flagged for closer examination, since perturbative corrections such as the perturbative triples (T) become less reliable as near-degeneracies develop. For sufficiently small systems, FCI may provide an additional reference within configured feasibility limits.

The methods assigned to each regime are configurable, and they define a practical policy for adaptive method selection rather than a general ranking of method accuracy. If diagnostic concerns arise primarily from numerical difficulties, the advisor can recommend further examination of the single-reference calculation instead of treating nonconvergence alone as justification for an active-space calculation. This can avoid unnecessary computational cost. Each recommendation records its supporting diagnostic evidence, allowing users to assess the basis for the proposed calculation. Once accepted, the recommendation is implemented as an explicit refinement calculation.

\noindent\textbf{Active-space selection.} When a CAS-based treatment is selected, CAFES determines which orbitals and how many electrons to include in the active space. Because this choice defines the electronic configurations treated explicitly, CAFES combines numerical diagnostics with chemical information to construct candidate active spaces. Orbital energies and occupations help identify potentially relevant orbitals, while chemical valence information and atomic valence active space (AVAS) projections relate the selection to target atomic orbitals~\cite{sayfutyarova2017automated}. For individual calculations and adaptive studies, CAFES can retain both a compact, chemically motivated candidate and an expanded candidate based on additional diagnostic evidence, allowing users to compare their orbital content and computational requirements.

To support this comparison, the \texttt{ActiveSpaceAudit} module displays the candidate orbitals, energies, occupations, and available atomic-orbital contributions, together with the selection rationale and checks on the electron and orbital counts for user approval. If orbital localization or ordering is applied, the corresponding transformations are also recorded, linking the reviewed selection to the orbital representation supplied to the solver. Users can therefore examine the numerical evidence and chemical assumptions underlying each candidate. Within a study, CAFES coordinates active-space choices across related tasks to maintain a consistent chemical definition of the active space.

\subsection{Post-processing}

When a study contains many tasks, locating and assembling the quantities needed for analysis can be time-consuming. CAFES uses agent assistance to extract these quantities from the recorded results and organize them for analysis, reducing manual effort.
CAFES presents the results in comparison tables and plots. For example, molecular energies can be plotted along a geometry scan, while double occupancy and spin correlations can be examined as functions of model interaction parameters. Line plots, scatter plots, and heatmaps support comparisons across varying parameters. For specialized analyses, users can register additional routines in the capability registry, extending the procedures available to the agent.

Interpreting these comparisons requires considering both the physical variations under study and any changes in computational treatment. CAFES therefore retains calculation settings and method changes in the study records so that their effects can be assessed alongside plotted trends. This context is particularly important when a plot shows an apparent discontinuity. The associated diagnostics, electronic-state assignments, and computational settings can help determine whether the discontinuity reflects a physical change and guide further analysis or refinement.

\section{Applications\label{sec:results}}

We demonstrate CAFES through three computational studies. In the first study, described in Section~\ref{sec:lutein}, we compute the ground- and excited-state energies of the xanthophyll lutein for varying geometry using DMRG--CASSCF simulations. Orbital preparation and MPS transfer are used to maintain consistent active-space definitions and reuse converged states across related tasks.
The second study, described in Section~\ref{sec:honeycomb}, explores the quantum phases of the honeycomb Hubbard model using DMET. The calculations are coupled to an adaptive sampling procedure that uses computed observables to select additional interaction parameters for evaluation.
In the third study, described in Section~\ref{sec:dataset}, CAFES generates a quantum-chemical dataset with CCSD labels by combining registered molecular-dynamics and configuration-selection modules. Together, these studies show how CAFES supports different research workflows while reusing common capabilities for validation, execution, and data management.

\begin{figure*}[t]
    \centering
    \includegraphics[width=0.9\linewidth]{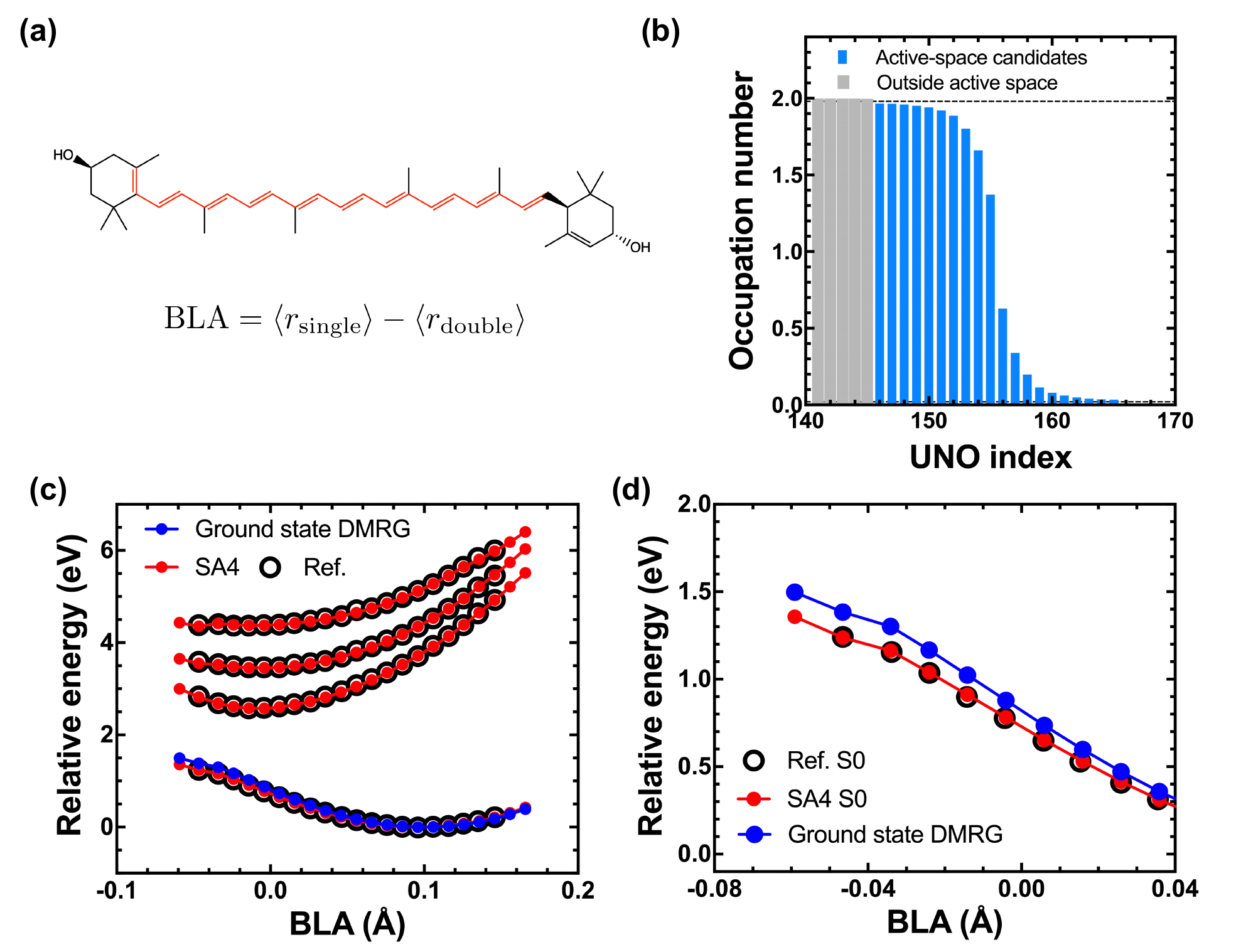}
    \caption{Study of the low-lying electronic states of xanthophyll lutein using DMRG-CASSCF. (a) Molecular structure of lutein. (b) Active-space selection. (c) Relative energies from Ref.~\citep{gianni2026lutein} (black circles), state-specific ground-state DMRG-CASSCF calculations (blue dots), and four-state-averaged DMRG-CASSCF calculations (SA4; red dots). (d) Enlarged view of the region from negative to small positive BLA values, highlighting the differences between the state-specific and SA4 results. 
    }
    \label{fig:lutein_results}
\end{figure*}

\subsection{Lutein: active-space selection and DMRG-CASSCF calculation\label{sec:lutein}}

Lutein (Fig.~\ref{fig:lutein_results}(a)) is a carotenoid pigment involved in light harvesting and photoprotection in photosynthetic complexes. Its extended $\pi$-conjugated backbone gives rise to low-lying electronic states with multiconfigurational character, making electron correlation important for describing their energies and ordering. Lutein therefore provides a molecular example for studying active-space selection and the structural dependence of strongly correlated states. Following Giann{\`\i} et al.~\cite{gianni2026lutein}, we characterize the backbone geometry using the bond-length alternation (BLA), defined as the difference between the mean lengths of alternating C--C bonds along the conjugated chain.

Varying the BLA provides a controlled way to examine the structural dependence of the low-lying states. Because multireference calculations are computationally demanding, this study also tests the reuse of converged results across related tasks. The geometries of neutral lutein ($\mathrm{C_{40}H_{56}O_2}$) were taken from the literature~\cite{gianni2026lutein}, and the calculations used the def2-SVPD basis, comprising 1,294 atomic orbitals (AOs). We performed state-specific DMRG-CASSCF calculations for the singlet ground state and equal-weight four-state-averaged DMRG-CASSCF (SA4-DMRG-CASSCF) calculations for the ground state and three low-lying singlet excited states. Density fitting was used for both the mean-field reference and CASSCF calculations with the def2-SVP-JKFIT auxiliary basis. Orbital optimization and DMRG calculations were performed using \textsc{PySCF} and \textsc{Block2}, respectively.

Active-space selection combined atomic valence active space (AVAS) analysis~\cite{sayfutyarova2017automated} with natural-orbital occupations. The active-space audit identified a broad chemically motivated candidate containing 228 electrons in 190 orbitals, which served as the starting point for further selection. The final CAS(20e,20o) was constructed from unrestricted natural orbitals obtained from the unrestricted Hartree--Fock density matrix. At the initial geometry, the selected orbitals had occupations ranging from 0.0347 to 1.9653 and formed ten complementary pairs with a total occupation of approximately two for each pair. The selection of ten such pairs is consistent with the 20 $\pi$ electrons associated with lutein's ten conjugated double bonds. Therefore, retaining these pairs allows electron redistribution between occupied-like and virtual-like orbitals within the conjugated chain. The active orbitals were localized using the Pipek--Mezey procedure~\cite{pipek1989fast} and reordered using Fiedler reordering~\cite{wouters2015chemps2}. This initial ordering was retained during orbital optimization.

CAFES maintained a consistent chemical definition of the active space and common electron and orbital counts across the study, while retaining the orbital indices and coefficient matrices specific to each calculation. Subsequent SA4 calculations were initialized from the converged orbitals and MPSs of previously completed geometries (Table S5). Transferring the orbitals together with the MPS preserves the orbital basis and ordering used to represent the correlated state. Each continuation was checked independently for convergence, and its settings, diagnostics, and restart data were retained.

Figure~\ref{fig:lutein_results} compares the state-specific ground-state and SA4 energy profiles with the literature reference data~\cite{gianni2026lutein}. The SA4 calculations used a bond dimension $M=2000$, with all molecular orbitals optimized. The SA4 energies are given relative to the SA4-S0 energy at BLA = 0.096 \AA, and the state-specific ground-state energies relative to the state-specific energy at the same geometry. With this choice, the two ground-state profiles nearly coincide near the minimum, whereas at negative BLA the state-specific energy increases more steeply than the SA4-$S_0$ profile. The excited states show a different structural dependence: the lowest sampled total energies of $S_1$--$S_3$ occur at $\mathrm{BLA}=-0.004$~\AA \ and increase toward larger positive BLA. Their low-energy regions are therefore shifted toward smaller BLA values relative to the ground state. Over the range covered by the literature data, the SA4 profiles closely follow the reference curves for both the ground and excited states, reproducing their overall shapes and the relative positions of their low-energy regions along the BLA coordinate.

\subsection{Honeycomb Hubbard model: adaptive parameter scans with DMET\label{sec:honeycomb}}

The honeycomb Hubbard model exhibits interaction-driven quantum phase transitions and competing correlated phases~\cite{meng2010quantum,sorella2012absence,chen2014honeycombDMET}. At half filling, competition between electron hopping and on-site repulsion produces semimetallic and antiferromagnetic (AFM) insulating regimes, while nearest-neighbor repulsion can favor charge-density-wave (CDW) order. Systematic scans of the interaction parameters are therefore needed to characterize its phase behavior, making the model a useful test case for an agentic framework.

The Hamiltonian used in this study is
\begin{equation}
\begin{aligned}
H={}&-t\sum_{\langle ij\rangle,\sigma}
\left(c_{i\sigma}^{\dagger}c_{j\sigma}+\mathrm{h.c.}\right)
+U\sum_i n_{i\uparrow}n_{i\downarrow}
+V\sum_{\langle ij\rangle,\sigma\sigma'}
n_{i\sigma}n_{j\sigma'} ,
\end{aligned}
\end{equation}
where $c_{i\sigma}^{\dagger}$ and $c_{i\sigma}$ are fermionic creation and annihilation operators for an electron with spin $\sigma\in\{\uparrow,\downarrow\}$ at site $i$, and $n_{i\sigma}=c_{i\sigma}^{\dagger}c_{i\sigma}$. $t$ is the nearest-neighbor hopping amplitude, and $U$ and $V$ are the on-site and nearest-neighbor Coulomb repulsion strengths, respectively. We use $t$ as the energy unit by setting $t=1$; thus, the values of $U$ and $V$ reported below correspond to the dimensionless ratios $U/t$ and $V/t$.

CAFES organizes the interaction-parameter scans using a shared lattice and embedding specification. Using the Model Hamiltonian Builder, we constructed a honeycomb lattice with $4\times4$ primitive cells and two sites per cell (Fig.~S1), giving 32 sites and 48 nearest-neighbor bonds under periodic boundary conditions. The lattice was half filled with 32 electrons, $(N_\uparrow,N_\downarrow)=(16,16)$. Its geometry, bond connectivity, boundary conditions, and electron count were fixed throughout the study, while $U$ and $V$ were varied.

Electron correlation was treated using DMET. The lattice was divided into eight non-overlapping impurities, each containing a $1\times2$ block of primitive cells and therefore 4 sites. Each impurity was coupled to a non-interacting bath and treated with ground-state FCI as the impurity solver. Separate scans were initialized from AFM and CDW unrestricted-HF guesses to sample competing broken-symmetry solutions. Details of the initialization, DMET convergence criteria, and numerical recovery are provided in the Supplementary Material. The mean-field, impurity, and embedding calculations used \textsc{PySCF} and \textsc{libDMET}.

Both scans began from the same $4\times4$ parameter grid, $U/t,\;V/t\in\{1,2,4,8\}$, which provided 16 initial points for each seed. We monitored the energy per site, double occupancy, and nearest-neighbor charge and spin correlations. The refinement observables and tolerances were specified through CAFES. The two scans were then refined independently using interpolation discrepancies in the energy per site and mean double occupancy. Additional points were sampled when either discrepancy exceeded its prescribed tolerance. Details of the interpolation, subdivision, and tolerances are provided in the Supplementary Material.

Figure~\ref{fig:honeycomb} compares the total energy, alternating double occupancy $\lvert D_{\mathrm{stag}}\rvert$, and nearest-neighbor charge and spin correlations for the AFM-initialized scan (a--d) and the CDW-initialized scan (e--h). Refinement concentrated additional calculations in the low-$V/t$ region and near rapid changes in the observables, while retaining coarser sampling over much of the high-$V/t$ region. CAFES therefore increased the sampling resolution where needed without uniformly refining the full parameter grid.

Both scans show similar trends: at small $V/t$ and sufficiently large $U/t$, $\lvert D_{\mathrm{stag}}\rvert$ is nearly zero and negative nearest-neighbor spin correlations indicate local antiferromagnetism. With increasing $V/t$, $\lvert D_{\mathrm{stag}}\rvert$ approaches $0.5$, spin correlations weaken, and connected charge correlations become small, consistent with charge disproportionation between the sublattices. At large $V/t$, the total energy increases with $U/t$, consistent with substantial double occupancy. Near the apparent phase boundary, however, the two initializations yield different charge and spin correlations: the CDW-seeded scan retains larger $\lvert D_{\mathrm{stag}}\rvert$ at lower $V/t$, whereas the AFM-seeded scan favors stronger antiferromagnetic correlations. This behavior suggests multiple self-consistent solutions associated with different local minima. Comparing their total energies helps identify the lower-energy solution, although further finite-size and convergence analyses may be needed to establish the favored phase.

\begin{figure*}[t]
    \centering
    \includegraphics[width=\linewidth]{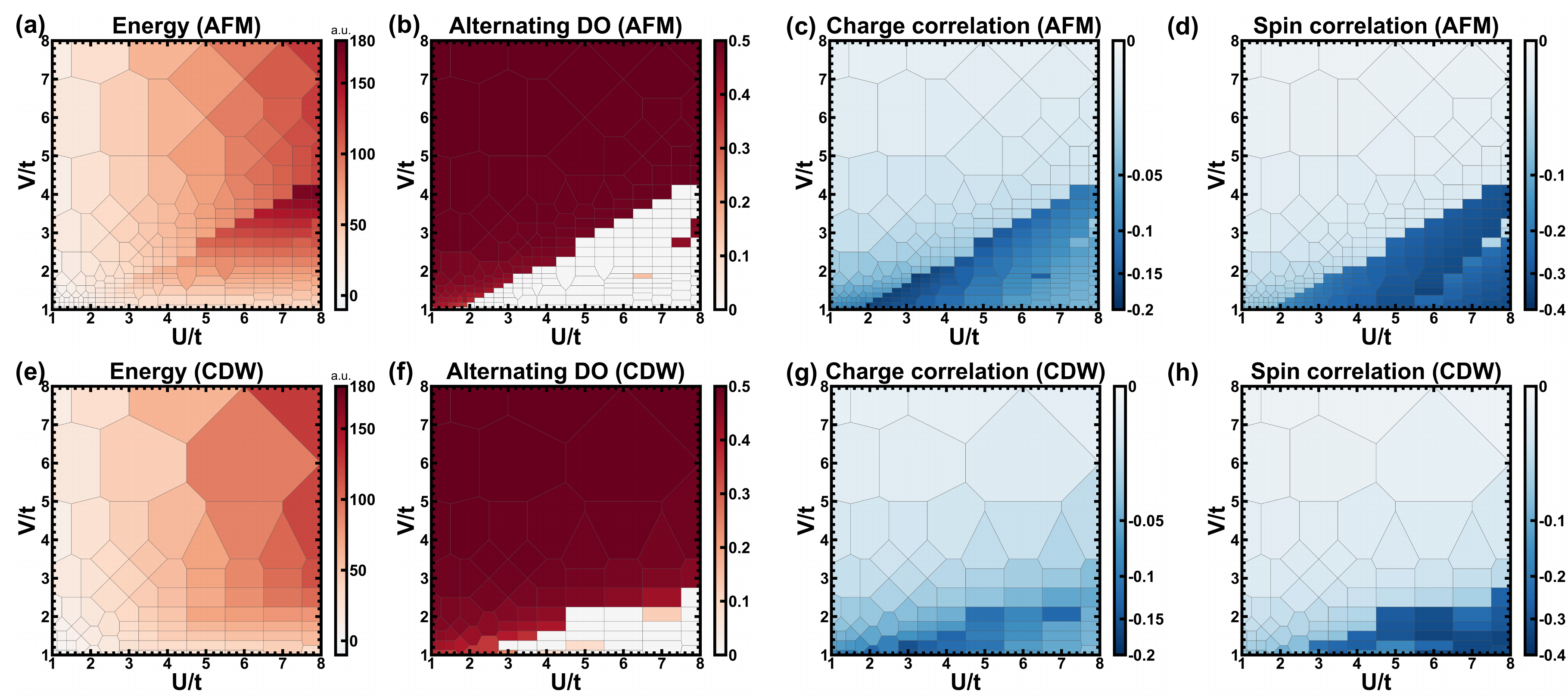}
    \caption{DMET interaction-parameter scans of the half-filled extended honeycomb Hubbard model, initialized with antiferromagnetic (AFM; top) and charge-density-wave (CDW; bottom) states. The AFM- and CDW-initialized scans comprise three and two adaptive refinement rounds, respectively.}
    \label{fig:honeycomb}
\end{figure*}

\subsection{Dataset generation: trajectory sampling and CCSD labeling\label{sec:dataset}}


\begin{figure*}[t]
    \centering
    \includegraphics[width=0.8\linewidth]{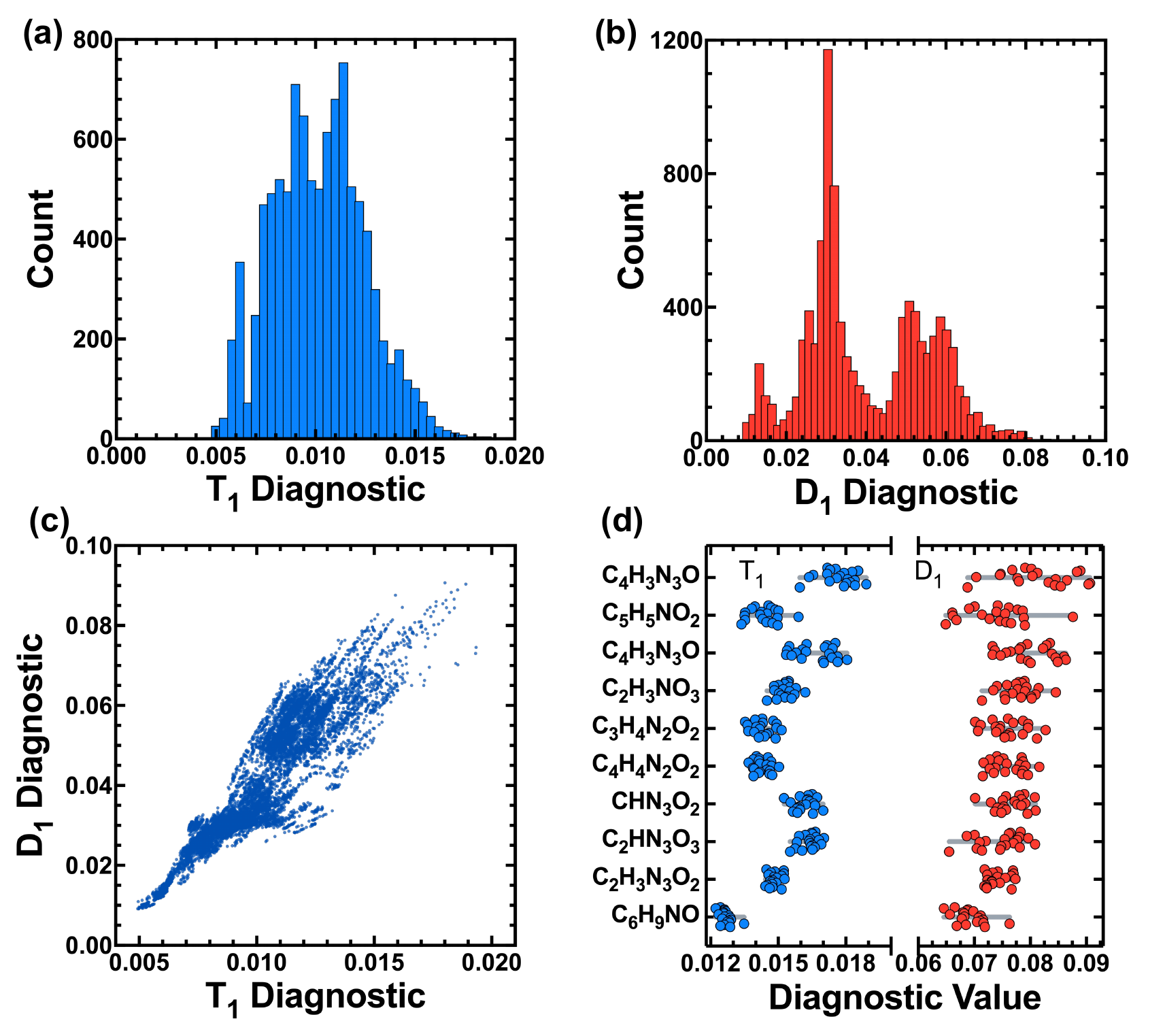}
    \caption{CCSD diagnostics for 10,000 configurations sampled from 500 QM9 molecules. (a,b) Distributions of the $T_1$ and $D_1$ diagnostics. (c) Joint distribution of $T_1$ and $D_1$. (d) Configuration-resolved $T_1$ (blue) and $D_1$ (red) values for the ten molecules showing the strongest diagnostic indications of multireference character.}
    \label{fig:Dataset}
\end{figure*}

A natural application of automated computational frameworks is the repeated generation of training data for machine learning. Most quantum-chemical datasets are based on DFT, but higher-accuracy data are important for improving models such as machine-learning interatomic potentials (MLIPs)~\cite{jacobs2025practical}. Correlated methods such as CCSD provide data for learning electronic properties beyond DFT and for comparing DFT and wavefunction-based descriptions of the same configurations. Because CCSD is substantially more expensive than DFT, an efficient workflow should combine broad DFT sampling with correlated calculations on a selected subset of configurations.

This procedure requires coordinating molecular dynamics, selecting configurations, and collecting the corresponding electronic-structure data. CAFES combines these steps within a single study, reducing the need for separate scripts and manual data movement. Following the functional and basis-set choices used in QH9~\cite{yu2023qh9}, we used CAFES to organize molecular-dynamics sampling and prepare configurations for subsequent CCSD calculations. The study included 500 QM9 molecules~\cite{ramakrishnan2014quantum}, with one \textit{ab initio} molecular-dynamics trajectory generated for each molecule using a common protocol. The DFT calculations used B3LYP/def2-SVP with an SCF energy-convergence threshold of $10^{-13}$~Ha~\cite{becke1993density}. Each trajectory was propagated in the microcanonical ensemble for 1~ps with a 1~fs timestep and Maxwell--Boltzmann initial velocities corresponding to 300~K, yielding 1001 configurations including the initial structure.

To obtain a manageable set of configurations while retaining the sampled potential-energy range, we introduced a post-processing module that selected snapshots according to their DFT potential energies. For each molecule, the 801 frames from 200 to 1000~fs were sorted by potential energy and divided into 20 groups with nearly equal populations. The lower-median frame from each group was retained, yielding 10,000 geometries across the 500 molecules. HF and CCSD calculations on these geometries completed the dataset, providing HF reference energies, overlap matrices, CCSD energies, and one-particle density matrices. The complete dataset contents are listed in Table S6. The dataset also includes the CCSD single-excitation amplitudes, $t_1$, together with the HF orbital coefficients, orbital energies, occupations, and indices, thereby specifying the orbital representation associated with the amplitudes.

To characterize the CCSD singles amplitudes across the sampled configurations, we evaluated the $T_1$ and $D_1$ diagnostics,
\begin{equation}
    \begin{aligned}
        T_1=\frac{\lVert t_1\rVert_F}{\sqrt{N_e}}, \quad
        D_1=\sigma_{\max}(t_1)
\end{aligned}
\end{equation}
where $N_e$ is the number of electrons, $\lVert\cdot\rVert_F$ denotes the Frobenius norm, and $\sigma_{\max}$ is the largest singular value of the singles-amplitude matrix.

Figures~\ref{fig:Dataset}(a,b) show the diagnostic distributions across the dataset. The $T_1$ values are concentrated around $0.01$, with most of the distribution lying approximately between $0.008$ and $0.014$. The $D_1$ distribution has a pronounced peak near $0.03$ and additional peaks around $0.05$--$0.06$. The joint distribution in Fig.~\ref{fig:Dataset}(c) shows an overall positive correlation, but configurations with similar $T_1$ values can have substantially different $D_1$ values. The electron-normalized total norm and the largest singular value therefore provide complementary information about the singles amplitudes. Evaluating both helps identify configurations for which the single-reference description warrants closer examination.

Figure~\ref{fig:Dataset}(d) highlights the ten molecules with the strongest indications of electron correlation in the dataset, as identified by the diagnostic assessment. Their diagnostic values also vary across the sampled geometries, with several molecules showing appreciable variation in $D_1$ despite relatively narrow $T_1$ distributions. For example, the selected $\mathrm{C_5H_5NO_2}$ molecule has $T_1$ values concentrated near $0.014$, while its $D_1$ values span approximately $0.065$--$0.09$. This variation illustrates the importance of evaluating both diagnostics for individual configurations rather than assigning a single correlation label to each molecule. The diagnostics identify potential limitations of the CCSD treatment, although they do not directly quantify errors in the resulting energies or density matrices.

This example illustrates how CAFES combines data generation, configuration selection, and diagnostic analysis within a common workflow. The diagnostics could help prioritize configurations for further validation, while the retained geometries and orbital information support calculation preparation and the CCSD results provide reference data for comparison. The workflow could also support data generation for reinforcement learning, allowing a learning agent to request additional electronic-structure calculations during training. These requests would follow the existing validation, execution, and reporting procedures, enabling the dataset to expand in response to changing training requirements.

\section{Discussion and Conclusion\label{sec:conclusion}}
We introduced CAFES, a correlation-aware framework for electronic-structure calculations built on the \textsc{PySCF} ecosystem. It supports molecular, crystalline, and model-Hamiltonian systems through a common study--task workflow that separates scientific planning from the execution of individual calculations. A capability registry defines the supported operations and their compatibility conditions, so workflow checks and compilation are performed by program logic.
Correlation diagnostics guide method selection and adaptive refinement, while active-space audits document the orbital information and chemical assumptions used in CAS-based calculations. Execution monitoring distinguishes scheduler completion from numerical convergence, and recovery procedures adjust numerical settings within predefined limits. We demonstrated CAFES through DMRG-CASSCF calculations for lutein, an adaptive DMET study of the honeycomb Hubbard model, and the generation of a CCSD-labeled molecular dataset. These examples cover molecular, lattice, and data-generation workflows using common procedures for validation, execution, refinement, and analysis.

The following sections discuss the scientific role, numerical reliability, and future outlook of agentic workflows in computational science.

\subsection{Scientific role of the agent}

Although many agentic frameworks focus on translating scientific problems into executable workflows, a more difficult challenge is deciding which calculations should be performed next based on evidence from previous results. This capability is particularly important for strongly correlated systems, where the appropriate method may vary across geometries or parameter regimes.

CAFES proposes refinements based on accumulated evidence, including observables and diagnostics from completed tasks, as demonstrated in the adaptive DMET study. These proposals remain subject to registry constraints, qualification checks, and human approval where required. By expressing different methods through common study and task specifications, CAFES also lowers the barrier to cross-method comparisons within a single study. The agent therefore acts more like a research assistant that interprets results, proposes and explains subsequent calculations, and supports scientific planning than like a workflow engine that executes a fixed sequence.

\subsection{Numerical reliability and human oversight}

A computational plan does not guarantee reliable results. Reliability requires assessing whether the requested procedure was followed, whether the numerical solution converged, and whether the method is suitable for the target properties. These questions require different evidence and cannot be reduced to a single task-success label.

In CAFES, registry-based qualification and workflow compilation verify that the requested procedure uses supported operations, while output inspection and numerical diagnostics assess convergence. Method suitability, however, is more difficult to automate: correlation diagnostics can reveal limitations of a single-reference treatment, but questions such as DMET convergence with respect to impurity size often require additional calculations and scientific judgment.

Researchers therefore retain responsibility for the scientific objective and consequential methodological decisions, including active-space selection and changes in the level of theory. Requiring approval for every refinement would limit scalability. Future versions could make review requirements depend on the diagnostic evidence, allowing calculations within validated regimes to proceed automatically while referring ambiguous cases to the researcher.

\subsection{Generality and future directions}

Molecular, crystalline, and model-Hamiltonian calculations require different diagnostics and physical interpretations, even when they share common needs for method comparison and adaptive refinement. CAFES accommodates these differences through system-specific fields and modules within a shared workflow structure. Adding a new method therefore requires not only a numerical implementation, but also a registry description of its interface and compatibility conditions and scientific guidance on its assumptions and limitations.

A further opportunity is to transfer computational experience between studies. Successful calculations, failures, and researcher corrections can inform later decisions when their physical and numerical context is retained. CAFES's curated knowledge layer provides a mechanism for recording and reviewing such information independently of individual conversations. This may allow future workflows to benefit from accumulated domain knowledge while keeping the assumptions and limitations of each recommendation visible.

Overall, CAFES provides a controlled framework for using agent assistance in strongly correlated electronic-structure studies. Its scientific value depends on the quality of the underlying methods, diagnostics, and validation procedures, but its explicit representations and execution records provide a foundation for reproducible, inspectable, and progressively more automated computational research.

\section*{Data Availability}

The CAFES source code and data generated for this work are available at
\url{https://github.com/lith0419/CAFES}. The CCSD-labeled QM9 dataset is available at
\url{https://doi.org/10.5281/zenodo.23068307}.

\section*{Acknowledgments}
This research is supported by a start-up fund from the Rutgers University.
We acknowledge the Office of Advanced Research Computing (OARC) at Rutgers University for computational resources.
We thank Laura Milena Pedraza Gonzalez for sharing structures of lutein, and Huanchen Zhai for useful discussions on Block2.

\section*{Supplementary Material}

Additional computational details, input specifications, convergence analyses, and supplementary figures are provided in the Supplementary Material.

\bibliographystyle{unsrtnat}
\bibliography{references}

\end{document}